\documentclass[aps,groupedaddress,%
amsmath,preprintnumbers,10pt,prl,twocolumn]{revtex4}

\usepackage{graphicx}
\usepackage{amsfonts}
\usepackage{amssymb}
\usepackage{amsbsy}
\usepackage{amsmath}

\def\p {\partial}

\def\be {\begin{equation}}
\def\ee  {\end{equation}}
\def\bea {\begin{eqnarray}}
\def\eea {\end{eqnarray}}
\def\nn {\nonumber}

\begin{document}
\preprint{ }
\title{Gravitational collapse of quantum matter}
\author{Benjamin K.  Tippett}
\email{ xxx@unb.ca}
\author{Viqar Husain}
\email{vhusain@unb.ca}

\affiliation{ Department of Mathematics and Statistics, University
of New Brunswick, Fredericton, NB, Canada E3B 5A3} 
\date{\today}

\begin{abstract}

 We describe a class of exactly soluble  models for gravitational collapse in spherical symmetry 
 obtained by patching  dynamical spherically symmetric exterior  spacetimes with  cosmological
 interior spacetimes.  These are generalizations of the Oppenheimer-Snyder type models
 to include classical and quantum scalar fields as sources for the interior metric, and null
fluids with pressure as sources for the exterior metric. In addition to dynamical exteriors, the models exhibit other novel features such as evaporating horizons and singularity avoidance without
quantum gravity. 
 
\end{abstract}

\maketitle

\section{Introduction}

Gravitational collapse of matter configurations has been an important area
of investigation for many years due mainly to the interest in understanding the dynamics
of black hole formation. There are only  a handful of analytical models in general relativity
beginning with the very first   by Oppenheimer and Schneider (OS) \cite{OS}. In their model stellar  matter is assumed to be homogeneous and isotropic and modelled by a closed FRW solution with pressureless dust. This was followed some years later by the Vaidya solution 
\cite{Vaidya} which describes the collapse of pressureless null dust to a Schwarzschild black hole. 
Models of this type have been much  studied  in various contexts \cite{lindquist1965,casadio1996,fayos1991,adler2005,fayos2008}
 
There is a generalization of the Vaidya  solution to a null dust with pressure, with equation of state  $P=k\rho$  \cite{H-Vaidya}, which gives the Reissner-Nordstrum charged black hole for $k=1$, and a more general class of hairy black holes for $k>1$ as the end point of collapse.   Although  these solutions are  analytic, they describe purely  ingoing (or outgoing) matter in spherical symmetry.  

The simplest asymptotically flat solutions with both inflow and outflow in spherical symmetry were found numerically, with a minimally coupled scalar field as the matter source \cite{choptuik}. These solutions exhibit a rich structure, including a discrete self-similarity and a mass scaling law for black holes.  This  work has led to much additional research on gravitational collapse \cite{Gundlach-rev}, including  quantum gravity inspired effects at the onset of black hole formation \cite{VH-qgcollapse, ZK-qgcollapse,pelota2009,modesto2004}. The latter are expected to play a fundamental role  in the late stages of collapse  and it is not unreasonable to expect that quantum gravity will drastically affect the strong field regime. Indeed these works  show that black holes form with a  mass gap, a result that was also suggested in an OS type model with a quantized scale factor in the interior region \cite{BGMS}.  The quantum gravity corrections used in the numerical simulations of scalar field collapse were inspired by the so-called polymer quantization procedure \cite{ashtekar2003} which grew out of the loop quantum gravity. 

An interesting feature of polymer quantization is that it introduces a length scale in addition to $\hbar$  into the quantum theory. This comes about because the choice of Hilbert space used to represent operators is such that  conventional momentum operators (ie. generators of translations) do not exist, but are defined indirectly as certain functions of translation operators. This is because
translation operators are not continuous in the translation parameter, unlike in Schrodinger/Heisenberg
quantization.  A direct consequence is that kinetic energy operators are bounded above. This may be viewed as an ultraviolet cutoff that is built in due to the choice of Hilbert space. It is this fact that has the potential to lead to interesting new physics at short distances while recovering standard quantum mechanics at large distances.  This quantization prescription  has been applied to the  scalar field \cite{HK,HHS-effec,HHS-prop,Mah} and to semiclassical gravity in the case of homogeneous cosmology \cite{HHS-cosm}. 

Motivated by these works we describe an analytical model of gravitational collapse that combines
three  features: (i) the basic OS idea, (ii) the generalization of the Vaidya metric \cite{H-Vaidya}  and (iii)  polymer quantization of the matter sector. The  last of these has been investigated in detail recently in the quantization of the scalar field on a Friedmann-Robertson-Walker (FRW) background, leading among other things, to the remarkable prediction of an extended inflationary period in the early universe without  a mass term or other scalar potential \cite{HHS-cosm}.  Our motivation for using this quantization is partly motivated by this result.  

The model we explore is a classical  FRW  interior metric sourced with a polymer quantized scalar field that is patched using junction conditions to the generalization of the Vaidya metric given in \cite{H-Vaidya}. The setting is thus that of a quantum field on a fixed classical background, and so does not incorporate any quantum gravity corrections (unlike Ref. \cite{BGMS} which has some of the same ingredients).  Our main result is that polymer quantization of matter is sufficient to  avoid a curvature singularity -- surprisingly without recourse to quantum gravity.

In the next section we describe the classical model. This is followed in section III by a discussion of the
polymer quantization of matter and its effect on the dynamics of the FRW scale factor. In Section IV we 
apply this dynamics to study collapse scenarios using the junction  conditions at the interface of the  
interior and exterior metrics. The concluding section contains a discussion of the main results and its  relation to some other  works.  
 
 \section{The model}
 
The collapse models  we consider are all  constructed by patching together an interior FRW spacetime with a generalized Vaidya-like exterior. This differs from the usual OS model in that the interior solution is to have quantized scalar field matter, and hence non-zero pressure. Therefore matching to a pressureless null dust  is not possible without surface stresses; this is why we must use a more general  exterior solution.  

The parametrization we use for the interior  metric $g_{a b}^-$ is 
\be 
ds_{-}^{2}=-dt^{2}+\frac{a^{2}(t)}{(1+r^{2}/4)^{2}}(dr^{2}+r^{2}d\theta^{2}+r^{2}\sin^{2}\theta d\phi^{2}),\label{eq:inmetric}
\ee
and that for  exterior metric  $g^+_{ab}$ is 
\be
ds_{+}^{2}=- f(v,R)\ dv^{2}+2 \ dvdR+R^{2}\left(d\theta^{2}+\sin^{2}\theta d\phi^{2}\right),
\label{eq:outmetric}
\ee 
which is written in advanced Eddington-Finkelstein coordinates $(v,r)$.  These coordinates are convenient because the trapping horizon, which is one of the objects of interest here, is given simply by  $f(v,R)=0$. We have taken the coordinates $(\theta,\phi)$ on the  $2-$spheres to be the same in  the two metrics. The coordinates $(t,r)$ of the interior metric and  $(v,R)$ of the exterior metric are of course different and so the matching of these metrics along a common  timelike $3-$surface 
$\Sigma$ must take this into account.  For this we need the induced metrics $h^\pm_{ab}$ and extrinsic curvatures $K^\pm_{ab}$ of $\Sigma$ from both sides to carry out the standard matching analysis.
  
\subsection{Matching surface: interior view}   
Let the metric $h_{ab}^-$ on $\Sigma$ from the FRW side be given by setting $r=r_0$, a constant. This is the natural choice that describes a $2-$sphere evolving along a timelike trajectory. The metric is  then 
\bea
ds_-^2 &=& h_{ab}^-\ dx^adx^b \nn\\
&=& -dt^{2}+\frac{a^{2}(t) r_0^2}{(1+r_0^{2}/4)^{2}}\left(d\theta^{2} + \sin^{2}\theta d\phi^{2}\right)\;.\label{inq} 
\eea
The unit timelike tangent and unit spacelike normal of  the surface $\Sigma$ are respectively   
\be
t^a_- = \left({\p\over \p t}\right)^a, \ \ \ \ n^a_- = {1\over a}\left(1+{r^2\over 4}\right) \left( \p\over \p r \right)^a
\ee
Using these, the non-zero components of the extrinsic curvature  
\be
K^-_{ab} = h^{-\ c}_{\ a} h^{-\ d}_{\ b}\  \nabla_{(c}n^-_{d)}
\ee
are 
\be
K_{tt}^-=0,\ \ \ \ \ \ K_\theta^{-\  \theta} =  K_\phi^{-\  \phi}=  {1-r_0^2/4\over a(t)\ r_0}.
\ee

\subsection{Matching surface: exterior view}

Let  the timelike surface from the exterior side be given by $R=R(t)$ and $v=v(t)$, where $t$ is the
interior's proper time coordinate.   The induced metric $h_{ab}^+$ on $\Sigma$ is then 
\bea
ds_+^2 &=& h_{ab}^+ dx^adx^b\nn\\
&=& -\left(f(v(t),R(t)) \dot{v}^2- 2\dot{R}\dot{v}  \right) dt^2 \nn\\&&+ R(t)^2 \left(d\theta^{2}+\sin^{2}\theta d\phi^{2}\right),
\label{outq}
\eea
where the dot denotes $d/dt$.

The unit timelike tangent  and unit spacelike normal of $\Sigma$ are respectively
\be
t^a_+ ={1\over  \sqrt{f\dot{v}^2 - 2 \dot{R} \dot{v} }}   \left[\dot{v} \left({\p  \over \p v}\right)^a 
+ \dot{R} \left(\p \over \p R  \right)^a \right], 
\ee
\be
n^+_a =  {1\over  \sqrt{f \dot{v}^2- 2 \dot{R} \dot{v}}} \left[-\dot{R}\ (dv)_a +  \dot{v}(dR)_a \right]. 
\ee 
The extrinsic curvature components 
\be
K^+_{ab} = h^{+\ c}_{\ a} h^{+\ d}_{\ b}\  \nabla_{(c}n^+_{d)}
\ee
are therefore 
\bea
K^+_{tt} &=& K^+_{ab}t^a_+t^b_+ \nn\\
&=& {\dot{v}^2 \left( f f_{,R}\dot{v} - f_{,v}\dot{v} - 3 f_{,R} \dot{R} \right) +4(\ddot{R}\dot{v} - \dot{R}\ddot{v})\over 2 \left(  f\dot{v}^2 - 2 \dot{R} \dot{v}   \right)^{3/2} }\nn \\
\eea
where $f_{,R} = \p f/\p R$ etc. and 
\be
K_\theta^{+\ \theta} =  K_\phi^{+\ \phi} =  {f\dot{v} - \dot{R} \over R \sqrt{f\dot{v}^2 - 2 \dot{R} \dot{v}}}
\ee

\subsection{Junction conditions}

The Israel junction conditions require continuity of the metric, and continuity of the extrinsic
curvature if there is to be no surface stress-energy.  Taken together these equations describe
the dynamics of the boundary surface. Matching the induced metric components gives
\bea 
&& a(t) b = R(t), \label{metric-m1}\\ 
&& f\dot{v}^2 -2\dot{R}\dot{v} =1\label{metric-m2},
\eea 
where we have set $b=r_0/(1+r_0^2/4)$, which is a constant.  
Matching the extrinsic curvature components, and using the previous equation  give  
\bea
&& f\dot{v} -\dot{R} =   c,
\label{Ktheta-m} \\
&&  \dot{v}^2\left( (f f_{,R} - f_{,v})\dot{v} - 3 f_{,R} \dot{R} \right)  \nn\\
&& +\ 2(\ddot{R}\dot{v} - \dot{R}\ddot{v})=0\label{Ktt-m}
\eea
where $c=(1-r_0^2/4)/ (1+r_0^2/4)$; note that $b^2 + c^2 =1$. The last condition  may be rewritten in a simpler
form by noting that the derivatives of (\ref{metric-m2}) and (\ref{Ktheta-m}) give
\be
2\dot{R}\ddot{v}  = -\dot{f}\dot{v}^2, \ \ \ \ \ \ \  2\dot{v}\ddot{R} = \dot{f}\dot{v}(2\dot{R} -f\dot{v}),
\ee
which leads to 
\be
K_{tt}^+ = - {f_{,v}\dot{v}^2\over 2\dot{R}}=0.
\label{Ktt-ms}
\ee

The four equations (\ref{metric-m1}-\ref{Ktheta-m}) and (\ref{Ktt-ms}) fully determine the dynamics of the boundary and the exterior metric function $f(R,v)$: a matter source for the interior FRW determines $a(t)$, and hence $R(t)$ through  eqn. (\ref{metric-m1}), the next two determine $f(\dot{R},c)$
and $\dot{v}(\dot{R},c)$, and the  last equation requires  $f=f(R)$ on the boundary. The second and third equations also give the  four-velocity of the boundary as seen from the 
exterior, ie. 
\bea
 \label{dots}
 {\bf t }^\pm&:=& (\dot{v}, \dot{R},0,0) \nn\\
 &=& \left( \frac{1}{f}\left(c\pm\sqrt{c^2-f}\right)  ,\pm\sqrt{c^2-f},  0,0\right).
 \eea
 The $\pm$ solutions correspond to collapsing and expanding solutions.  We note  also 
 that $K_{tt}^+=0$ implies that ${\bf t}^\pm $ are tangent to radial geodesics in the exterior spacetime.
 
We are interested in the function $R(v)$ which gives the trajectory of the FRW boundary. From  (\ref{dots}) this is 
\be
{dR\over dv} = {\pm f\sqrt{c^2-f}\over c\pm \sqrt{c^2-f}} = \pm\sqrt{c^2-f}\left( c\mp \sqrt{c^2-f}\right)
\label{star-trajectory}
\ee

This is equation we study  for various cases of interest  for the classical and quantum scalar field. Its use requires only the  function $f(R)$ obtained from the junction condition, which in turn depends on the
interior matter Hamiltonian. If the interior metric is taken to be flat FRW the matching conditions above remain valid with the changes $b=r_0$ and $c=1$. We restrict attention to this latter case for explicit
calculations and comment on the other values $c<1$ in the discussion section. 

We will see that for scalar field matter in the interior, the exterior metric function determined on the boundary, $f(R)$, has a {\it dynamical} extension which can be analytically determined and interpreted as a null fluid with pressure. Furthermore, if the interior matter field is quantized using
polymer quantization, which is a semiclassical approximation, the surface trajectory exhibits qualitatively new features. Both classical and quantum matter cases exhibit interesting dynamical horizon behaviour.

\section{Classical matter solutions}

Our goal is to construct models of gravitational collapse that go beyond the OS solution
by using quantum matter in the interior.  One way to do this is to begin with the  Hamiltonian formulation
of the gravity-matter dynamics in the interior and then quantize the matter canonical variables. This is a canonical approach to the semiclassical approximation. 

We begin with the  Arnowitt-Deser-Misner (ADM) canonical form of  the 3+1 action for Einstein gravity minimally coupled to a massless scalar field
\be
S= \int dt\int_\Sigma d^3x \left(\tilde{\pi}^{ab} \dot{q}_{ab}  + p_\phi\dot{\phi}- N{\cal H} - N^a{\cal C}_a\right),
\label{action}
\ee
where $(\tilde{\pi}^{ab}, q_{ab})$ are the ADM canonically conjugate variables and ${\cal H}$ and
${\cal C}_a$ are the Hamiltonian and diffeomorphism constraints, and $(\phi,p_\phi)$ are the scalar field canonical variables. Reduction to homogeneous and isotropic cosmology is attained by the parametrization
\be
q_{ab} = a^2(t) e_{ab},\ \ \ \ \ \ \ \ \  \tilde{\pi}^{ab}= {p(t)\over 2a} e^{ab} \sqrt{e},
\ee 
where  $e_{ab}$ is the flat spatial metric in Eqn. (\ref{eq:inmetric}).  
This gives the reduced action 
\be
S= V_0\int dt\left[ p \dot a  +p_\phi\dot \phi - N\left( -{p^2\over 24a} + {8\pi\over a^3}p_\phi^2  \right)\right] \\
\ee
for  flat FRW in the interior (which corresponds to $c=1$ and $b=r_0$ in the junction conditions), and 
$V_0 =\int d^3x$  comes from the reduction to homogeneity. 
 
\subsection{Dust: Oppenheimer-Snyder solution}
 
From a Hamiltonian perspective this model may be viewed as arising from the dynamics of the
canonically conjugate pair $(a,p)$. Evolution is given by  the Hamiltonian constraint above but with the
scalar field energy density term replaced by a constant:  
\be
{\cal H} \equiv-{p^2\over 24a} + 16\pi \rho_0 =0. 
\label{OS-cons}
\ee
where $\rho_0 = $ constant. The canonical equations of motion for lapse $N=1$ are 
\bea
\dot{a} &=&  \{a, {\cal H} \}=  -{p\over 12 a} \\
\dot{p} &=& \{p, {\cal H}\} = -\frac{p^2}{24a^2}.
\eea 
Rewriting the constraint equation (\ref{OS-cons}) using the $\dot{a}$ evolution equation, the junction condition 
$R(t) = r_0a(t)$ and the $\dot{R}$ equation (\ref{dots}) (with $c=1$) gives 
\be
16\pi \rho_0 = 6a\dot{a}^2 = \frac{6R}{r_0^3}(1-f(R)), 
\ee
which gives  the known result 
\be
f(R) = 1-\frac{2M}{R}, \quad  M= \frac{4\pi}{3}r_0^3 \rho_0,
\ee
which constrains the exterior metric to be Schwarzschild and provides the interpretation that the interior
is dust ball of radius $r_0$ and density $\rho_0$.
 
\subsection{Scalar field}

As a warm up to the quantum matter problem, it is useful to consider the interior FRW sourced with a
massless minimally coupled scalar field. Although this is a natural generalization, it doesn't appear to exist in the literature. As we will see the exterior metric takes an unusual form, but is one of a class of known solutions. 

Hamilton's equations obtained from the action (\ref{action}) with $N=1$ are  
\bea 
\dot{a} &=&  -{p\over 12 a}, \label{adot}\\
\dot{p} &=&  - {p^2\over 24 a^2} + {24\pi V_0p_\phi^2\over a^4},\\
\dot{p}_\phi&=&0 \\ 
\dot{\phi} &=& {16\pi V_0p_\phi\over a^3},
\eea
together with the Hamiltonian constraint 
\be
\mathcal{H}\equiv  -\frac{p^2}{24a} + \frac{8\pi V_0}{a^3}p_\phi^2 =0.
\label{Hc}
\ee 

The procedure for finding the exterior metric function on the boundary uses the Hamiltonian
constraint, the equation of motion for the scale factor, the junction condition   $R(t)=ba(t)$,
and the trajectory equation  (\ref{dots}). We have 
\bea 
\label{f-proc}
 \mathcal{H} (p_\phi,a,p) &=& \mathcal{H} (p_\phi,a,\dot{a}) 
 =\mathcal{H} (p_\phi,R,\dot{R}) \nn\\
 &=&\mathcal{H} (p_\phi,R, f(R))=0.
\eea
For the Hamiltonian constraint  (\ref{Hc}) this gives 
\bea
 8\pi V_0 p_\phi^2 &=&  6\left(R\over r_0\right)^4 \left( {\dot{R}^2\over r_0^2}\right) \nn\\
 &=&    
  6\left(R\over r_0\right)^4 \left( {1 -f (R)\over r_0^2}\right).
\eea
 This determines the exterior metric function on the boundary to be   
 \be
   f(v,R)|_\Sigma  = 1-   {4\pi V_0p_\phi^2\over 3} {r_0^6 \over R^4} \equiv 1- {\alpha\over R^4},
   \label{boundary-fn}
 \ee
where $\alpha \ge 0$ is a constant since $p_\phi$ is a constant of the motion. This suggests unusual exterior spacetimes as extensions of this boundary function.  
 

\begin{figure*} 
  \label{k=1c}
  \begin{center} 
  \includegraphics[width=6in]{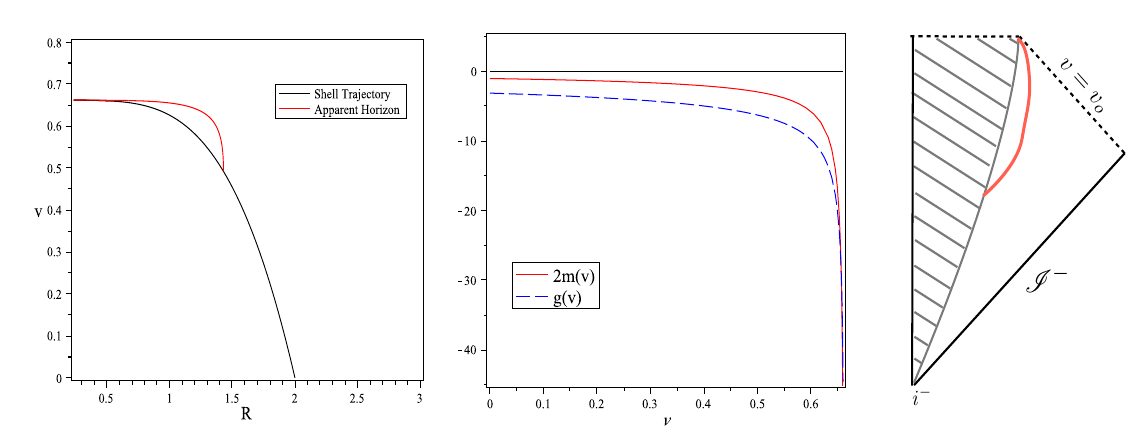} 
 \caption{$k=1$ exterior with classical scalar field interior:  the exterior spacetime is dynamical  and has null singularity  at a fixed value of advanced time (here $v_0\sim0.67$). } 
 \end{center}
 \end{figure*}
 
 
 
 \begin{figure*} 
  \label{k=2c}
  \begin{center} 
  \includegraphics[width=6in]{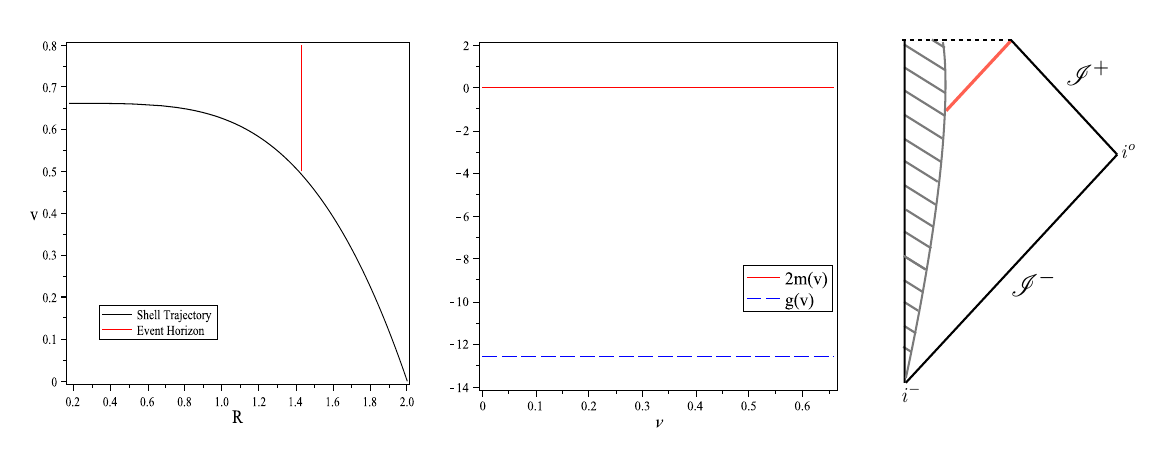} 
  \caption{$k=2$ exterior with classical scalar field interior:  the only solution is a static exterior (constant $m(v)$ and $g(v)$). This is qualitatively similar to the Oppenheimer-Snyder solution. }
  \end{center}
 \end{figure*}
 
 
 \subsection{Exterior solution}

It is interesting that there is  a  large class of exact solutions with  null fluid stress energy tensor  \cite{H-Vaidya} that provide possible extensions to the exterior of  the boundary function (\ref{boundary-fn}). These are given by  the  metric
 \be
  ds^2 = -\left(1- \frac{2M(R,v)}{R} \right) dv^2 \nn\\
 + 2dvdR + R^2 d\Omega^2,
 \label{fluid-metric}
 \ee
which  arises from the source
 \bea
 T_{ab} &=& {1\over 2\pi R^2}  {\p M\over \p v}\ v_av_b \nn\\
 &&+  \rho(v,R) w_{(a}v_{b)} +   P(v,R) \left(g_{ab} + w_{(a}v_{b)}\right).\nn \\
 \label{Tab}
 \eea
The pressure $P(v,R)$ and energy density $\rho(v,R)$ are given by  
\be
P = k {g(v)\over 4\pi r^{2k+2}} = k\rho,
\ee
where
\be
M(R,v) = m(v) - \frac{g(v)}{2(2k-1)R^{2k-1}} 
\ee
and
\bea
v_a&=&(1,0,0,0),\nn\\
 w_a&=&(F/2,-1,0,0),
\eea
are the future pointing null vectors; $F$  is the coefficient of $dv^2$ in the metric, and
$k$ is a real parameter.

From the form of the metric (\ref{fluid-metric}) it is apparent that there is a unique value, $k=2$, that 
extends the boundary function  $f(R)= 1-\alpha/R^4$ derived in  eqn. (\ref{boundary-fn}) to a static
exterior. For this case we must have  $m(v)=0$ and $g(v)=$ constant.   

There are however interesting dynamical possibilities for the exterior for other values of $k$. 
 These are obtained  by finding the functions $m(v)$ and $g(v)$ such that on the interface
$R(v)$ given by (\ref{star-trajectory}) we have 
\be
1- {2m(v)\over R(v)} + {g(v)\over (2k-1)R(v)^{2k} }= 1- {\alpha\over R(v)^4}.
\ee
and 
\be
{2\dot{m}(v) \over R(v)} = {\dot{g}(v) \over (2k-1)R(v)^{2k}}.
\ee
The first of these is the requirement that the exterior metric match the boundary function, and the
second  is the junction condition (\ref{Ktt-ms}).  Taking the derivative of the first  and using the second
gives the unique one parameter ($k$) family of solutions
\bea 
m(v) &=&   {\alpha(2k-4)\over (2k-1)R(v)^3},\\
g(v) &=& -3\alpha R(v)^{2k-4}.
\eea
 The case $k=2$ gives $m(v)=0$ and $g(v)=-3\alpha$, which is the static  solution already noted above. 
All other values of $k$ provide a dynamical exterior. 

Typical classical matter solutions are exhibited in  the Figures 1-3. Each shows the junction  trajectory, the metric functions $m(v)$ and $g(v)$ and the corresponding Penrose diagram with a (dynamical) horizon trajectory. All calculations  are with the initial condition $R(v=0)=2$ and the parameter  values $p_\phi=1$ and $r_0=1$.

\noindent\underbar{$k=1$}:  (Fig. 1) A dynamical horizon forms and evolves on a time like trajectory and then becomes null and tangent to the surface trajectory; there is a  curvature singularity  at the corresponding value of $v$. The horizon starts to shrink because the metric functions are such that an ingoing flux of positive energy evolves to an ingoing flux of negative energy. This is apparent from examining the stress-energy tensor (\ref{Tab}). The null singularity is naked because timelike observers
in the exterior region intersect the $v=v_0$ line in finite proper time.

 \noindent\underbar{$k=2$}: (Fig. 2)  This is the static case and is qualitatively similar to  the Oppenheimer-Snyder solution: $m(v)=0$ and $g(v) =$ constant $< 0$ which gives a single horizon. 
The Penrose diagram is therefore the  standard one representing gravitational collapse to a black hole.

 \noindent\underbar{$k=3$}: (Fig. 3) This is another case with dynamical exterior and has the surprising feature that the dynamical horizon becomes null at finite $v$ and extends to spatial infinity. The  change from $k=1$ to $k=3$ dramatically changes   the horizon dynamics, as may be expected since the order
 of the equation $F(r,v)=0$ is quite different.  The metric functions are such that the null energy condition
 holds. 
 
 
 \begin{figure*} 
  \label{k=3c}
  \begin{center} 
  \includegraphics[width=6in]{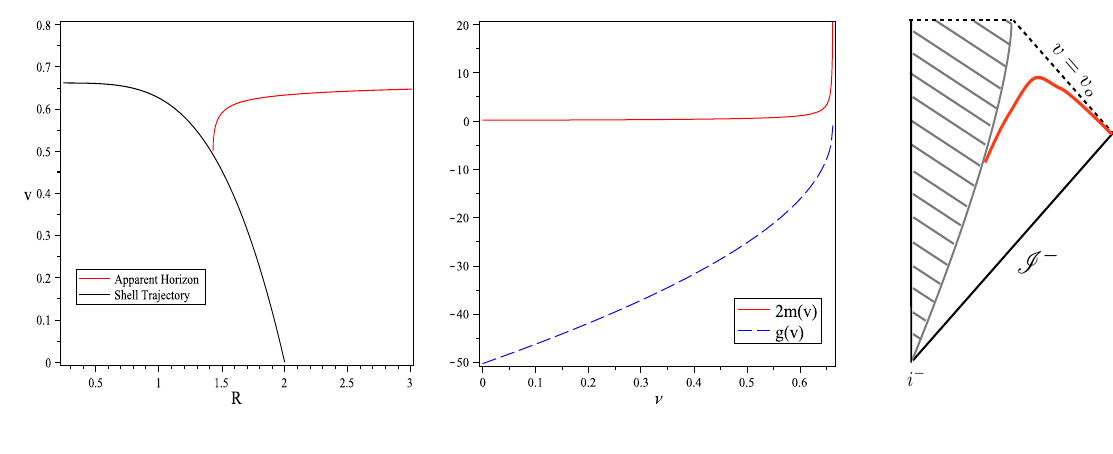} 
  \caption{$k=3$ exterior with classical scalar field  interior: the exterior spacetime is dynamical and has a null singularity at a value of advanced time (here $v_0\sim 0.67$). The apparent horizon extends to spatial infinity.}
  \end{center}
 \end{figure*}


\section{Quantum matter solutions }


\begin{figure*} 
  \label{k=2q}
  \begin{center} 
  \includegraphics[width=6in]{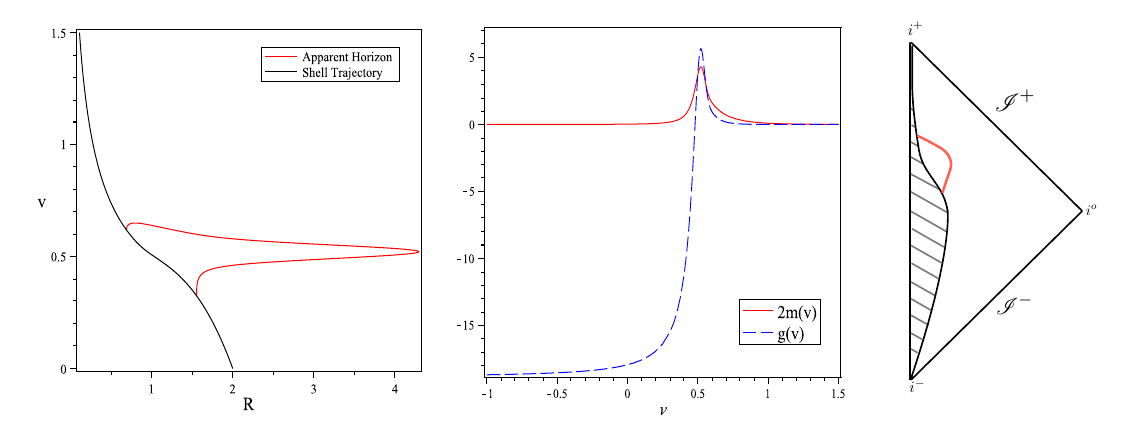} 
  \caption{$k=2$ exterior with quantum scalar field interior:  the shell trajectory exponentially approaches
 $r=0$ and there is no curvature singularity. The apparent horizon forms and evaporates in finite time.}
 \end{center}
 \end{figure*}


In this section we consider the same model as in the last section, but with a quantum 
scalar field. The resulting model is in the context of the usual semiclassical approximation where the expectation value of quantized  matter is used as a source for  the classical gravitational field.
 
In this approach the interior dynamics  is derived from the semiclassical constraint 
\begin{equation}\label{SCHam}
    {\cal H}_{sm}=  -\frac{p^2}{24a} + \frac{8\pi V_0}{a^3}\langle \psi | \widehat{p^2_\phi} | \psi \rangle =0
\end{equation}
where $|\psi\rangle$ is suitably chosen state could depend on the scale factor $a(t)$ and other 
parameters. A guiding principle for selecting the  state is that the resulting spacetime give the expected classical result for a large universe, but with increasing  quantum affects  as it shrinks in size
and approaches the would be classical singularity. This is the type of state we select, and it does depend on the scale factor. 
 
The derivation of the boundary function is identical to that for the classical case since the junction
conditions do not depend on the nature of the interior dynamics.  The semiclassical constraint gives the formula
\be
f(v,R)|_\Sigma =  1- \frac{4\pi V_0 r_0^6}{3R^4}\ \langle \psi | \widehat{p^2_\phi} | \psi \rangle(R),
\label{Q-boundfn}
\ee
where the scale factor (and hence $R$) dependence of the expectation value is made explicit. As we will see, it is this dependence which can drastically modify the exterior solution. 
 
 The quantization procedure we follow is that  described in \cite{HHS-cosm},where a particular type of polymer quantization is applied to the scalar field. The variables used in this approach are motivated by but different from the ones that arise from loop quantum gravity  \cite{ashtekar2003}. The essential
 difference is the manner in which the field translation operator is defined. 
 
In the following we review this quantization procedure and summarize how it leads
to a modified interior solution. The main feature of interest  is that the interior scalar field
energy density turns out to be bounded above; matching this  new interior solution to the exterior
solution (\ref{fluid-metric}) gives an evaporating horizon and a remnant as the end point  of gravitational collapse. This is one of the main results of this paper.

 \subsection{Polymer quantization}

This quantization  method does not use the standard canonical variables $(\phi, P_\phi)$
as the starting point of quantization, but rather the new variables 
 \begin{equation}\label{BasicVariables}
\phi_f \equiv \int d^3x \sqrt{q} \, f(\textbf{x}) \phi(\textbf{x}) ,
\quad U_{\lambda} \equiv \exp\left( \frac{i \lambda
p_{\phi}}{\sqrt{q}} \right),
\end{equation}
where the smearing function $f(\textbf{x})$ is a scalar \cite{HHS-cosm}. The parameter $\lambda$ 
is a spacetime constant with dimensions of $(\text{mass})^{-2}$, and the $\sqrt{q}$ factor in the 
exponent is required to balance the density weight of $p_\phi$.  These  variables satisfy the
Poisson algebra
\begin{equation}
\label{basic-pb} \{\phi_f,U_\lambda\} = if\lambda U_\lambda.
\end{equation}

Specializing to a scalar field on an FRW background, these variables become
\begin{equation}\label{ReducedVariables}
\phi_f = V_0 a^3 \phi , \quad U_\lambda= \exp\left(i\lambda
p_{\phi}/a^3 \right),
\end{equation}
where $V_0 = \int d^3 x$ is a fiducial comoving volume and $f$ is set to unity since 
this is a reduction of $\phi_f$ to the spatially homogeneous case.  
The Poisson bracket of the reduced variables is the  same as that of the
unreduced ones (\ref{basic-pb}).

 Quantization proceeds by realizing the Poisson algebra
(\ref{basic-pb}) as a commutator algebra on a suitable Hilbert
space;  the choice for polymer quantization has the basis
$\{|\mu\rangle | \mu \in \mathbb{R}\}$ with inner product
\begin{equation}\label{InnerProduct}
\langle\mu^\prime |\mu\rangle = \delta_{\mu,\mu^\prime},
\end{equation}
where $\delta$ is the generalization of the Kronecker delta to the
real numbers. The operators $\hat\phi_f$ and $\hat
U_\lambda$ have the action
\begin{equation}\label{operator actions}
    \hat\phi_f |\mu\rangle = \mu|\mu\rangle,
    \quad \hat U_{\lambda} |\mu\rangle = |\mu + \lambda\rangle;
\end{equation}
i.e., $|\mu\rangle$ is an eigenstate of the smeared field
operator $\hat\phi_f$, and $\hat U_\lambda$ is the generator of
field translation.

With this realization it is evident that configuration eigenstates
are normalizable. This is one of the main difference between the
polymer and Schr\"odinger quantization schemes. It is because of
this that the momentum operator does not exist in this quantization,
but must be defined indirectly using the translation generators by
the relation
\begin{equation}\label{SFMomentum}
\widehat{p_{\phi}^\lambda}= \frac{a^3}{2i\lambda} (\hat{U}_{\lambda } -
\hat{U}^{\dagger}_{\lambda }) ~.
\end{equation}

\subsection{Quantum energy density}

To compute the expectation value in the semiclassical constraint (\ref{SCHam})
we use the  Gaussian coherent state peaked at the phase
space values $(\phi_0,P_\phi)$: 
\bea
    |\psi\rangle &=& \frac{1}{\mathcal{N}} \sum_{k=-\infty}^{\infty}   c_k
    |\lambda_k\rangle,\nn\\
    c_k &\equiv& e^{-(\phi_k-\phi_0)^2/2\sigma^2} e^{-iP_\phi \phi_k V_0},
\eea
where $\phi_k = \lambda_k/V_0 a^3$ is an eigenvalue of the scalar
field operator derived from $\hat{\phi}_f$ in Eq.~(\ref{operator
actions}).  This state is chosen because it ensures that a large universe satisfies
the classical Einstein equations, but introduces quantum corrections when the universe
is sufficiently small \cite{HHS-cosm}. 

Computing the expectation value of the operator $ \widehat{p^2_\phi}$ in the above state we find
\be
\langle\psi |  \widehat{p^2_\phi} | \psi \rangle =
\frac {a^6}{2\lambda^{2}} \left[1-\exp(-\Theta^2/\Sigma^2)\cos(2 \Theta)\right], 
\label{eq:polymerH}
\ee
where 
\be
 \Theta \equiv \lambda P_\phi a^{-3}, \quad \Sigma \equiv \sigma
    V_0 P_\phi,
 \ee
 are scale invariant variables; the first is just the exponent in the field translation operator, and the second is a measure of the with of the semi-classical state (since $p_\phi$ is a constant of also of the semiclassical equations of motion if the scalar field potential is zero). Hence the metric function on the boundary given by Eqn. (\ref{Q-boundfn}) is   
 \be
 f(R) = 1- \frac{2\pi V_0 R^2}{3\lambda^2} \left[ 1-\exp(-\Theta^2/\Sigma^2)\cos(2 \Theta) \right]_{a=R/r_0}.
 \ee
 It is straightforward to verify that in the limit $\lambda\rightarrow \infty$  
 this reduces to (\ref{boundary-fn}). The same is true for large $R$ with fixed lambda. 
 However for sufficiently small $R$ the exponential factor goes to zero and $f(R)$ is regular as $R\rightarrow 0$. This of course  is just a consequence of the behaviour of the scale factor with the quantum matter source in the interior (which is communicated to the exterior  via the junction conditions).  It is this  fact that significantly modifies the late stages of collapse in this model and gives singularity
 avoidance. 

A numerical integration of the star trajectory appears in Fig. 4 for the case $k=2$. This is to be compared with Fig. 2 which is the same situation but with classical scalar field. Some salient features are that the junction trajectory goes approaches $R=0$ zero exponentially indicating a rapid disappearance of mass rapid  together with horizon formation and evaporation.  Correspondingly, the metric function $m(v)$ indicates  a non-zero mass function that rises and falls to zero, and a variation of $g(v)$ from negative to positive and eventually to zero.  The weak energy condition is violated for  the range of $v$ values in which the apparent horizon shrinks \cite{H-Vaidya}.

\section{Conclusions and Discussion}

We have constructed a new class of exactly solvable gravitational collapse models with two main results. The first is a classical generalization of the OS models to include scalar field matter together with a  clear interpretation of the exterior metric. A novel feature  of this is that the  exterior spacetime can be static or dynamic depending on the value of the  equation of state parameter $k$ in the stress-energy tensor. 

The second is a result in semiclassical gravity using the polymer quantization method, where we find that the behaviour of classical models is drastically modified due to matter quantization. Specifically for the $k=2$ case we find that a singularity inside a horizon is replaced by  formation and subsequent evaporation of  an apparent horizon, with an exponentially disappearing  mass.  Although our
model does not have all the features of a full field theoretical collapse model, such as the scalar field in spherical symmetry, it is interesting that it gives  physically desirable features of gravitational collapse that have been conjectured by various authors.  

The semiclassical model also provides a scenario to explore other quantum states, values of $k$, and 
the case of closed and hyperbolic interior metrics. An additional feature is that the junction  equation (\ref{dots})  permits both expanding and contracting trajectories that permit a matching of 
these solutions.  An example where this is possible arises in the closed universe case  for which  the parameter $c$ in (\ref{dots}) must be less than one. This means that the trajectory reaches a turning point at a finite advanced time value (unlike the flat case discussed here).  This makes  it possible to construct solutions that  join collapsing and expanding branches at points where $dR/dv =0$; at such points  ${\bf t}^\pm = (1,0,0,0)$  and the two trajectories have the same acceleration $d^2R/dv^2$. This may provide an exactly solvable model  where a star forms, becomes a  black hole, and then evaporates. 

\bigskip

  We thank Andreas Kreienbuehl and Sanjeev Seahra for discussions. 
This work supported by Natural Science and Engineering Research Council of Canada.  

\bibliographystyle{apsrev}
\bibliography{frw-vaidya}

\begin{thebibliography}{21}
\expandafter\ifx\csname natexlab\endcsname\relax\def\natexlab#1{#1}\fi
\expandafter\ifx\csname bibnamefont\endcsname\relax
  \def\bibnamefont#1{#1}\fi
\expandafter\ifx\csname bibfnamefont\endcsname\relax
  \def\bibfnamefont#1{#1}\fi
\expandafter\ifx\csname citenamefont\endcsname\relax
  \def\citenamefont#1{#1}\fi
\expandafter\ifx\csname url\endcsname\relax
  \def\url#1{\texttt{#1}}\fi
\expandafter\ifx\csname urlprefix\endcsname\relax\def\urlprefix{URL }\fi
\providecommand{\bibinfo}[2]{#2}
\providecommand{\eprint}[2][]{\url{#2}}

\bibitem[{\citenamefont{Oppenheimer and Snyder}(1939)}]{OS}
\bibinfo{author}{\bibfnamefont{J.~R.} \bibnamefont{Oppenheimer}}
  \bibnamefont{and} \bibinfo{author}{\bibfnamefont{H.}~\bibnamefont{Snyder}},
  \bibinfo{journal}{Physical Review} \textbf{\bibinfo{volume}{56}},
  \bibinfo{pages}{455} (\bibinfo{year}{1939}).

\bibitem[{\citenamefont{Vaidya}(1951)}]{Vaidya}
\bibinfo{author}{\bibfnamefont{P.~C.} \bibnamefont{Vaidya}},
  \bibinfo{journal}{Phys. Rev.} \textbf{\bibinfo{volume}{83}},
  \bibinfo{pages}{10} (\bibinfo{year}{1951}).

\bibitem[{\citenamefont{Lindquist et~al.}(1965)\citenamefont{Lindquist,
  Schwartz, and Misner}}]{lindquist1965}
\bibinfo{author}{\bibfnamefont{W.}~\bibnamefont{Lindquist}, \bibfnamefont{R.}},
  \bibinfo{author}{\bibfnamefont{R.~A.} \bibnamefont{Schwartz}},
  \bibnamefont{and} \bibinfo{author}{\bibfnamefont{W.}~\bibnamefont{Misner},
  \bibfnamefont{C.}}, \bibinfo{journal}{Physical Review}
  \textbf{\bibinfo{volume}{137}}, \bibinfo{pages}{1364} (\bibinfo{year}{1965}).

\bibitem[{\citenamefont{Casadio and Venturi}(1996)}]{casadio1996}
\bibinfo{author}{\bibfnamefont{R.}~\bibnamefont{Casadio}} \bibnamefont{and}
  \bibinfo{author}{\bibfnamefont{G.}~\bibnamefont{Venturi}},
  \bibinfo{journal}{Classical and Quantum Gravity}
  \textbf{\bibinfo{volume}{13}}, \bibinfo{pages}{2715} (\bibinfo{year}{1996}).

\bibitem[{\citenamefont{Fayos et~al.}(1991)\citenamefont{Fayos, Ja{\'e}n,
  Llanta, and Senovilla}}]{fayos1991}
\bibinfo{author}{\bibfnamefont{F.}~\bibnamefont{Fayos}},
  \bibinfo{author}{\bibfnamefont{X.}~\bibnamefont{Ja{\'e}n}},
  \bibinfo{author}{\bibfnamefont{E.}~\bibnamefont{Llanta}}, \bibnamefont{and}
  \bibinfo{author}{\bibfnamefont{J.~M.~M.} \bibnamefont{Senovilla}},
  \bibinfo{journal}{Classical and Quantum Gravity}
  \textbf{\bibinfo{volume}{8}}, \bibinfo{pages}{2057} (\bibinfo{year}{1991}).

\bibitem[{\citenamefont{Adler et~al.}(2005)\citenamefont{Adler, Bjorken, Chen,
  and Liu}}]{adler2005}
\bibinfo{author}{\bibfnamefont{R.~J.} \bibnamefont{Adler}},
  \bibinfo{author}{\bibfnamefont{J.~D.} \bibnamefont{Bjorken}},
  \bibinfo{author}{\bibfnamefont{P.}~\bibnamefont{Chen}}, \bibnamefont{and}
  \bibinfo{author}{\bibfnamefont{J.~S.} \bibnamefont{Liu}},
  \bibinfo{journal}{American Journal of Physics} \textbf{\bibinfo{volume}{73}},
  \bibinfo{pages}{1148} (\bibinfo{year}{2005}).

\bibitem[{\citenamefont{Fayos and Torres}(2008)}]{fayos2008}
\bibinfo{author}{\bibfnamefont{F.}~\bibnamefont{Fayos}} \bibnamefont{and}
  \bibinfo{author}{\bibfnamefont{R.}~\bibnamefont{Torres}},
  \bibinfo{journal}{Classical and Quantum Gravity}
  \textbf{\bibinfo{volume}{25}}, \bibinfo{pages}{175009}
  (\bibinfo{year}{2008}).

\bibitem[{\citenamefont{Husain}(1996)}]{H-Vaidya}
\bibinfo{author}{\bibfnamefont{V.}~\bibnamefont{Husain}},
  \bibinfo{journal}{Physical Review} \textbf{\bibinfo{volume}{D53}},
  \bibinfo{pages}{1759} (\bibinfo{year}{1996}), \eprint{gr-qc/9511011}.

\bibitem[{\citenamefont{Choptuik}(1993)}]{choptuik}
\bibinfo{author}{\bibfnamefont{M.~W.} \bibnamefont{Choptuik}},
  \bibinfo{journal}{Phys. Rev. Lett.} \textbf{\bibinfo{volume}{70}},
  \bibinfo{pages}{9} (\bibinfo{year}{1993}).

\bibitem[{\citenamefont{Gundlach}(2003)}]{Gundlach-rev}
\bibinfo{author}{\bibfnamefont{C.}~\bibnamefont{Gundlach}},
  \bibinfo{journal}{Phys. Rept.} \textbf{\bibinfo{volume}{376}},
  \bibinfo{pages}{339} (\bibinfo{year}{2003}), \eprint{gr-qc/0210101}.

\bibitem[{\citenamefont{Husain}(2008)}]{VH-qgcollapse}
\bibinfo{author}{\bibfnamefont{V.}~\bibnamefont{Husain}},
  \bibinfo{journal}{Advanced Phys. Letters}  (\bibinfo{year}{2008}),
  \eprint{0801.1317}.

\bibitem[{\citenamefont{Ziprick and Kunstatter}(2009)}]{ZK-qgcollapse}
\bibinfo{author}{\bibfnamefont{J.}~\bibnamefont{Ziprick}} \bibnamefont{and}
  \bibinfo{author}{\bibfnamefont{G.}~\bibnamefont{Kunstatter}},
  \bibinfo{journal}{Phys. Rev.} \textbf{\bibinfo{volume}{D80}},
  \bibinfo{pages}{024032} (\bibinfo{year}{2009}), \eprint{0902.3224}.

\bibitem[{\citenamefont{Modesto}(2004)}]{modesto2004}
\bibinfo{author}{\bibfnamefont{L.}~\bibnamefont{Modesto}},
  \bibinfo{journal}{Physical Review D} \textbf{\bibinfo{volume}{70}},
  \bibinfo{pages}{124009} (\bibinfo{year}{2004}).

\bibitem[{\citenamefont{Pelota and Kunstatter}(2009)}]{pelota2009}
\bibinfo{author}{\bibfnamefont{A.}~\bibnamefont{Pelota}} \bibnamefont{and}
  \bibinfo{author}{\bibfnamefont{G.}~\bibnamefont{Kunstatter}},
  \bibinfo{journal}{Physical Review D} \textbf{\bibinfo{volume}{80}},
  \bibinfo{pages}{044031} (\bibinfo{year}{2009}).

\bibitem[{\citenamefont{Bojowald et~al.}(2005)\citenamefont{Bojowald, Goswami,
  Maartens, and Singh}}]{BGMS}
\bibinfo{author}{\bibfnamefont{M.}~\bibnamefont{Bojowald}},
  \bibinfo{author}{\bibfnamefont{R.}~\bibnamefont{Goswami}},
  \bibinfo{author}{\bibfnamefont{R.}~\bibnamefont{Maartens}}, \bibnamefont{and}
  \bibinfo{author}{\bibfnamefont{P.}~\bibnamefont{Singh}},
  \bibinfo{journal}{Phys. Rev. Lett.} \textbf{\bibinfo{volume}{95}},
  \bibinfo{pages}{091302} (\bibinfo{year}{2005}), \eprint{gr-qc/0503041}.

\bibitem[{\citenamefont{Ashtekar et~al.}(2003)\citenamefont{Ashtekar,
  Fairhurst, and Willis}}]{ashtekar2003}
\bibinfo{author}{\bibfnamefont{A.}~\bibnamefont{Ashtekar}},
  \bibinfo{author}{\bibfnamefont{S.}~\bibnamefont{Fairhurst}},
  \bibnamefont{and} \bibinfo{author}{\bibfnamefont{L.}~\bibnamefont{Willis},
  \bibfnamefont{J.}}, \bibinfo{journal}{Classical and Quantum Gravity}
  \textbf{\bibinfo{volume}{20}}, \bibinfo{pages}{1031} (\bibinfo{year}{2003}).

\bibitem[{\citenamefont{Hossain et~al.}(2009)\citenamefont{Hossain, Husain, and
  Seahra}}]{HHS-effec}
\bibinfo{author}{\bibfnamefont{G.~M.} \bibnamefont{Hossain}},
  \bibinfo{author}{\bibfnamefont{V.}~\bibnamefont{Husain}}, \bibnamefont{and}
  \bibinfo{author}{\bibfnamefont{S.~S.} \bibnamefont{Seahra}},
  \bibinfo{journal}{Phys. Rev.} \textbf{\bibinfo{volume}{D80}},
  \bibinfo{pages}{044018} (\bibinfo{year}{2009}), \eprint{0906.4046}.

\bibitem[{\citenamefont{Laddha and Varadarajan}(2010)}]{Mah}
\bibinfo{author}{\bibfnamefont{A.}~\bibnamefont{Laddha}} \bibnamefont{and}
  \bibinfo{author}{\bibfnamefont{M.}~\bibnamefont{Varadarajan}},
  \bibinfo{journal}{Class. Quant. Grav.} \textbf{\bibinfo{volume}{27}},
  \bibinfo{pages}{175010} (\bibinfo{year}{2010}), \eprint{1001.3505}.

\bibitem[{\citenamefont{Hossain
  et~al.}(2010{\natexlab{a}})\citenamefont{Hossain, Husain, and
  Seahra}}]{HHS-prop}
\bibinfo{author}{\bibfnamefont{G.~M.} \bibnamefont{Hossain}},
  \bibinfo{author}{\bibfnamefont{V.}~\bibnamefont{Husain}}, \bibnamefont{and}
  \bibinfo{author}{\bibfnamefont{S.~S.} \bibnamefont{Seahra}},
  \bibinfo{journal}{Phys. Rev.} \textbf{\bibinfo{volume}{D82}},
  \bibinfo{pages}{124032} (\bibinfo{year}{2010}{\natexlab{a}}),
  \eprint{1007.5500}.

\bibitem[{\citenamefont{Husain and Kreienbuehl}(2010)}]{HK}
\bibinfo{author}{\bibfnamefont{V.}~\bibnamefont{Husain}} \bibnamefont{and}
  \bibinfo{author}{\bibfnamefont{A.}~\bibnamefont{Kreienbuehl}},
  \bibinfo{journal}{Phys. Rev.} \textbf{\bibinfo{volume}{D81}},
  \bibinfo{pages}{084043} (\bibinfo{year}{2010}), \eprint{1002.0138}.

\bibitem[{\citenamefont{Hossain
  et~al.}(2010{\natexlab{b}})\citenamefont{Hossain, Husain, and
  Seahra}}]{HHS-cosm}
\bibinfo{author}{\bibfnamefont{G.~M.} \bibnamefont{Hossain}},
  \bibinfo{author}{\bibfnamefont{V.}~\bibnamefont{Husain}}, \bibnamefont{and}
  \bibinfo{author}{\bibfnamefont{S.~S.} \bibnamefont{Seahra}},
  \bibinfo{journal}{Phys. Rev.} \textbf{\bibinfo{volume}{D81}},
  \bibinfo{pages}{024005} (\bibinfo{year}{2010}{\natexlab{b}}),
  \eprint{0906.2798}.

\end{thebibliography}

\end{document}